\documentclass[10pt, twocolumn]{article} 

\usepackage[utf8]{inputenc} 

\usepackage{geometry} 
\usepackage{graphicx} 

\usepackage{titling}
\usepackage[explicit]{titlesec}
\usepackage{blindtext}
\usepackage[normalem]{ulem}
\usepackage{lastpage}
\usepackage[usenames,dvipsnames]{color}

\usepackage{booktabs} 
\usepackage{subfig} 
\usepackage{url}
\usepackage{amsmath}
\usepackage{abstract}
\usepackage{upgreek}
\usepackage{multirow}
\usepackage{units}

\definecolor{darkgray}{gray}{0.40}

\usepackage{fancyhdr} 
\newcommand{\published}[1]{\gdef\puB{#1}}
\newcommand{\puB}{}

\pretitle{\begin{center}}
\posttitle{\par\end{center}}

\preauthor{\begin{center}}
\postauthor{\par\end{center}}

\DeclareRobustCommand{\authorblock}{
\begin{center}
\textbf{Daniel Hampf\textsuperscript{\#}} \\
daniel.hampf@dlr.de \\ 
\vspace{1 em}
Paul Wagner\textsuperscript{\#}\\
paul.wagner@dlr.de \\ 
\vspace{1 em}
Wolfgang Riede\textsuperscript{\#} \\
wolfgang.riede@dlr.de \\ 
\vspace{1 em}

\textsuperscript{\#}Institute of Technical Physics, German Aerospace Center, Stuttgart, Germany \\ 

\end{center}}

\renewcommand{\thesection}{\MakeUppercase{\roman{section}.}}
\titleformat{\section}[hang]{\normalfont\filcenter}{\thesection}{1em}{\uppercase{\uline{#1}}}

\renewcommand{\thesubsection}{\MakeUppercase{\roman{section}.\roman{subsection}.}}
\titleformat{\subsection}{\normalfont}{\thesubsection}{0pt}{\uline{#1}}
\titleformat{\subsubsection}{\normalfont}{}{1 em}{\uline{#1}}

\published{IAC-14,A6.P,70x24899}
\title{\MakeUppercase{Optical technologies for observation of low earth orbit objects}}
\author{\authorblock}
\date{}

\begin{document}

\twocolumn[
\maketitle

\thispagestyle{fancy}

\vspace{-15mm}
\begin{onecolabstract}
In order to avoid collisions with space debris, the near Earth orbit must be continuously scanned by either ground- or spaced-based facilities. For the low Earth orbit, radar telescopes are the workhorse for this task, especially due to their continuous availability. However, optical observation methods can deliver complementary information, especially towards high accuracy measurements.

Passive-optical observations are inexpensive and can yield very precise information about the apparent position of the object in the sky via comparison with background stars. However, the object's distance from the observer is not readily accessible, which constitutes a major drawback of this approach for the precise calculation of the orbital elements.

Two experimental methods have been devised to overcome this problem: Using two observatories a few kilometres apart, strictly simultaneous observations of the same object yield an accurate, instantaneous 3D position determination through measurement of the parallax. If only one observatory is available, a pulsed laser can be used in addition to the passive-optical channel to measure the distance to the object, in a similar fashion as used by the satellite laser ranging community. However, compared to conventional laser ranging, a stronger laser and more elaborate tracking algorithms are necessary. The two approaches can also be combined by illuminating the object with a pulsed laser from one observatory and measuring the return times at both observatories.

These techniques are explored by German Aerospace Center in Stuttgart using its orbital debris research observatory, in cooperation with the Satellite Laser Ranging station in Graz and the Geodetic Observatory in Wettzell. This contribution will present some of the results and plans for further measurement campaigns.
\end{onecolabstract}

\vspace{1cm}
]

\begin{table*}[t]
\centering
\begin{tabular}{ lrrr}
   \hline                     
						& \bf{FLI ProLine 16803}	& \bf{Andor iXon Ultra} 	& \bf{Andor Zyla}		\\
   Field of view [$^\circ$]			& 0.7 $\times$ 0.7		& 0.16 $\times$ 0.16 & 0.32 $\times$ 0.27	\\
   Resolution [pixel]				& 4096 $\times$ 4096	& 512 $\times$ 512	& 2560 $\times$ 2160	\\
   Pixel size [$\upmu \text{m}$]		& 9  				& 16			& 6.5				\\
   Scale [arcsec / pixel]			& 0.63 			& 1.12		& 0.45			\\
   Integration time [s]			& 0.02 - 3600		& 0.01 - 30		& $2.5 \times 10^{-5}$ - 30	\\
   Readout time [s]				& 2.1	 			& $\sim$ 0		& 0.01			\\
   Readout noise [e$^-$ / pix]		& 9				& 0.4 (equiv.)	& 2.6				\\
   Dark noise [e$^-$ / pix / s]		& 0.07			& 0.001		& 0.14			\\
   Peak quantum efficiency [\%]		& 50				& 90			& 60				\\
   \hline  
 \end{tabular}
 \caption{Specifications of the three cameras used for space debris observations. Field of view and scale are valid in combination with the Planewave CDK 17 telescope. \cite{fli, andor, andor_zyla, jonas}.}
 \label{specifications_cameras}
\end{table*}

\section{Introduction}
Today, about 17,000 objects of size over $\unit[10]{cm}$ are catalogued in low Earth orbit (LEO). Of these, less than 1,000 are active space-craft, the rest being different sorts of space debris: Upper stages, defunct satellites, and fragmented parts \cite{quarterlynews2014i1}. Most of them cluster at altitudes around 500 to $\unit[1000]{km}$: At lower altitudes the residual atmosphere decelerates the object substantially, which causes an atmospheric re-entry within a few years, while higher altitudes are less used by satellites operators in the first place. The objects move with velocities of about $\unit[8]{km/s}$, hence even particles in the centimetre range exhibit kinetic energies that can destroy an active payload.

While efforts are made to reduce the amount of debris produced during routine operations, e.g.\ by reducing the payload's altitude towards the end of its lifetime \cite{UNO_space_debris_mitigation}, the amount of debris increases constantly. The main sources of new debris are fragmentation of larger objects, partly induced by collisions with other debris objects, new launches, and objects released during routine operations.

Options to remove objects from orbit are under investigation \cite{2013EUCAS...4..735L}. Large debris objects may be removed by dedicated vehicles, which approach the debris object, capture it by appropriate means and drag it into the atmosphere \cite{2013JGCD...36..743M}. Smaller particles may be removed by ground-based facilities, using a very powerful laser to decelerate the object \cite{2014AcAau..93..418P}. Both options are expensive and not yet available. 

Currently, the most effective way to deal with space debris are an efficient space surveillance system coupled with a conjunction analysis, which allows spacecraft operators to initiate collision avoidance manoeuvres in time. The three main benchmarks of any orbital debris observation system are its accuracy, its capability to observe small objects, and the time needed per observation. From an organisational point of view, the building and operating costs have to be taken into account as well. Since each object's orbital parameters must be updated regularly, usually every few days at least, and to remedy visibility limitations, a network of observation stations is needed. Such a network may include monitoring systems of different type, cost, and capabilities for specific tasks (e.g.\ cheaper low resolution systems for initial object detection, high precision systems for subsequent measurement of orbital parameters).

Today, almost all observations of objects in LEO are conducted by radar systems. However, optical observations can be complimentary to radar observations and may work in cooperation with other systems in an efficient network. Their main drawback is the dependence on weather conditions, while their advantages are high accuracy, high sensitivity and low cost. Generally, optical observations of LEO objects can be passive or active. In passive mode, only the apparent two-dimensional position in the sky is measured, while active observations using laser ranging additionally yield the distance to the object and thus a three-dimensional position. Both options are under investigation by the German Aerospace Center at the orbital debris research observatory in Stuttgart, Germany. This contribution  shows recent results and lines out further plans.

\section{The orbital debris research observatory}
\begin{figure}[t]
   \centering
   \includegraphics[width=\columnwidth]{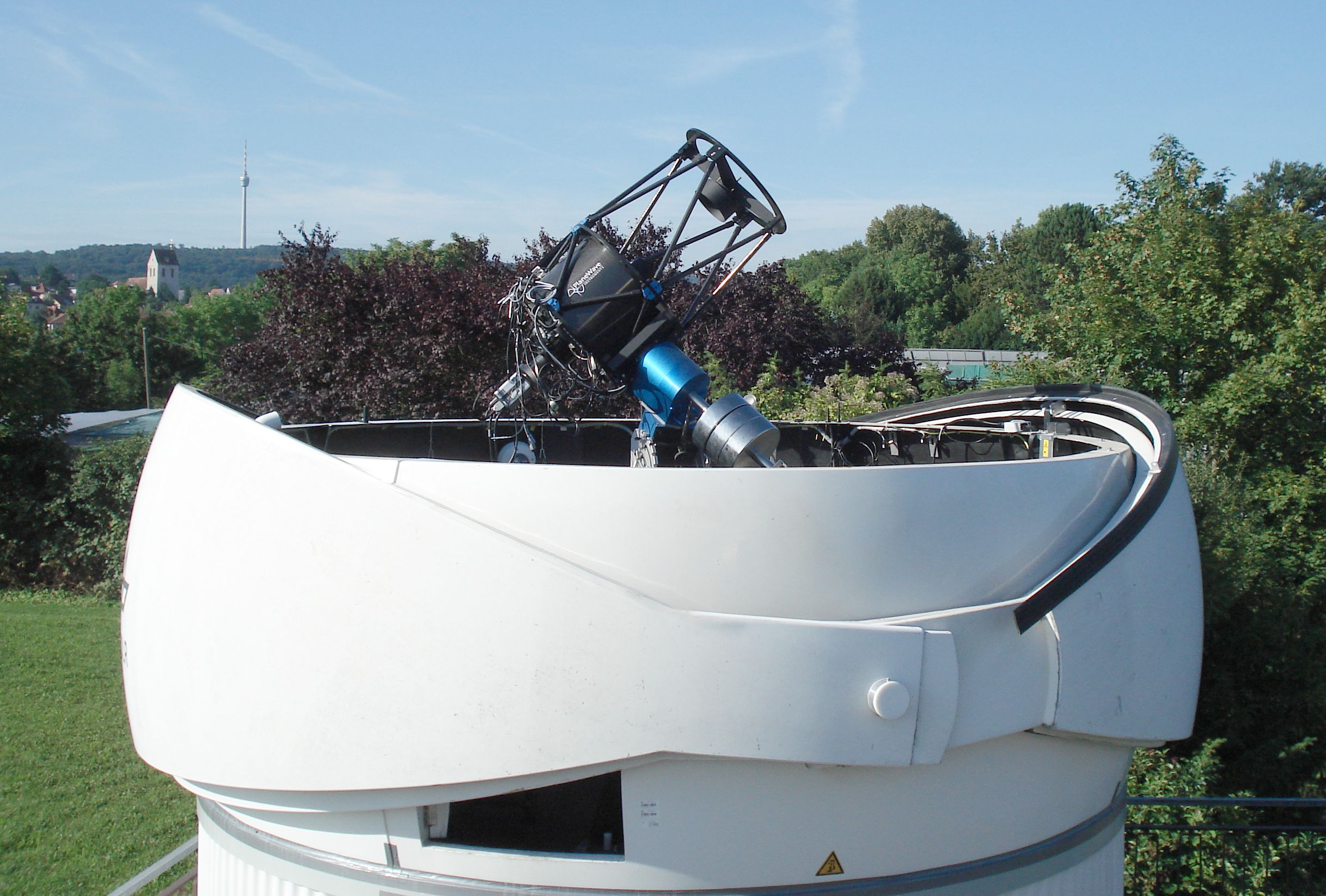}
   \caption{Orbital debris research observatory in Stuttgart, Germany.}
   \label{UFO}
 \end{figure}

The orbital debris research observatory \cite{Hampf2013} is situated on the grounds of Stuttgart's historical astronomical observatory at 48,78239 N, 9,19642 E and at an altitude of $\unit[350]{m}$ (see Figure \ref{UFO}). It is equipped with an Astelco NTM-500 direct-drive equatorial mount which offers a positioning velocity of up to $20^\circ / s$ and a tracking accuracy of about two arcseconds when tracking LEO objects, using a closed loop control. As main telescope, a Planewave CDK 17 with an aperture of $\unit[43]{cm}$ and a focal length of $\unit[294]{cm}$ is used.

 Several cameras are available for use at the main telescope (see table \ref{specifications_cameras}). For routine operations, the Andor Zyla sCMOS camera has been found to offer the best performance in terms of sensitivity, resolution and timing. For highest sensitivity, the Andor iXon emCCD camera can be used, albeit at the price of a reduced resolution. For highest resolution and largest field of view, the FLI ProLine Astro-CCD camera is used.

Since the exact time of exposure is important for many measurements, a GPS-based event timer has been developed, which can measure the UTC time of a TTL signal with an accuracy of better than $\unit[100]{\upmu s}$. In this time span, a typical LEO object moves by less than a metre, therefore this accuracy is sufficient. For both Andor cameras, the TTL signal is specified to occur within $\unit[25]{ns}$ relative to the actual exposure, while the timing resolution of the FLI camera is in the order of a few $\unit[10]{ms}$. Therefore, only the Andor cameras are used for time critical measurements. All cameras can also be triggered externally to start an exposure exactly (within their timing accuracy) on the start of a full second, which is very useful for simultaneous observations with two cameras.

For additional observations, the cameras can be used piggybacked on the main telescope or at a separate mount. Several smaller telescopes and photo-lenses are available for wide field-of-view observations. An ASA DDM60 PRO portable direct drive mount is available for bi-static measurements.

Studies on passive-optical debris observations are conducted regularly since April 2013 (see section \ref{passive}). Currently, the research observatory is being upgraded with an active (laser ranging) channel (see section \ref{active}).

\section{Passive optical measurements}
\label{passive}
\subsection{Position measurements}
The telescope can be used to track any object with known orbital elements, e.g\ in TLE format. The tracking can be continuous or in leap-frog mode. The former variant is mainly relevant for the planned active-optical measurements (see section \ref{active}). In the latter case, the telescope is positioned at the object's trajectory, a few seconds ahead of its current position, and put into sidereal tracking. At the appointed time, the camera starts an exposure which shows the stars as points and the target object as arc or track. In such a picture, the object's position at start and end of the exposure can be evaluated with high accuracy (usually better than 5 pixels) relative to the star positions. Using star catalogues, the sky coordinates of the target object can usually be determined with an accuracy of better than a few arcseconds, using an astronomic plate solving algorithm\footnote{DC-3 Dreams Pinpoint: \url{http://pinpoint.dc3.com/}}.

\begin{figure}[t]
   \centering
   \includegraphics[width=\columnwidth]{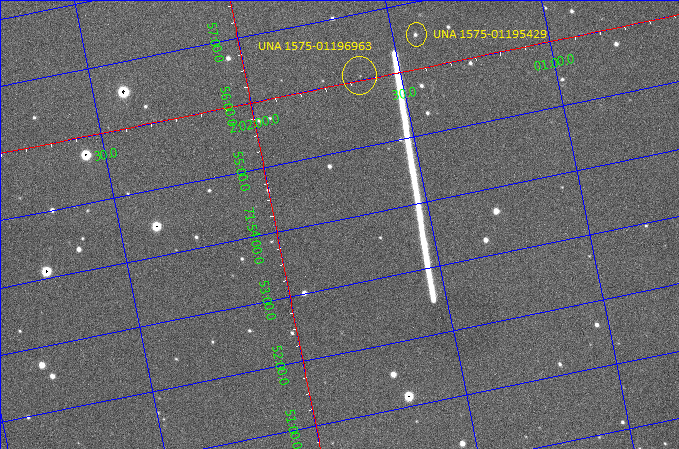}
   \caption{Image recorded of Envisat on May 18$^\mathrm{th}$, 2014, 20:30:37 UTC. The coordinate lines have been added during plate solving. Two stars have been marked for manual cross check with a star map of the area. The exposure time was 0.3 s.}
   \label{envisat}
 \end{figure}

Figure \ref{envisat} shows  a picture taken in this mode. The upper end of the track, which corresponds to the start of the exposure, is positioned at RA = 2$^h$ 1$^{min}$ $31^s$, dec= 71$^\circ$ 56$^{min}$ $19^s$. A set of such images can be used to derive the cross-track and along-track deviation of the object from its trajectory prediction and to update the orbital parameters. Comparisons with concurrent SLR measurements have shown an absolute angular accuracy in the range of ten arcseconds (about $\unit[50]{m}$ in $\unit[1000]{km}$ distance). A further improvement on this is envisaged, e.g.\ by improvement of the image analysis routines.

\subsection{Light curves}

\begin{figure}[t]
   \centering
   \includegraphics[width=\columnwidth]{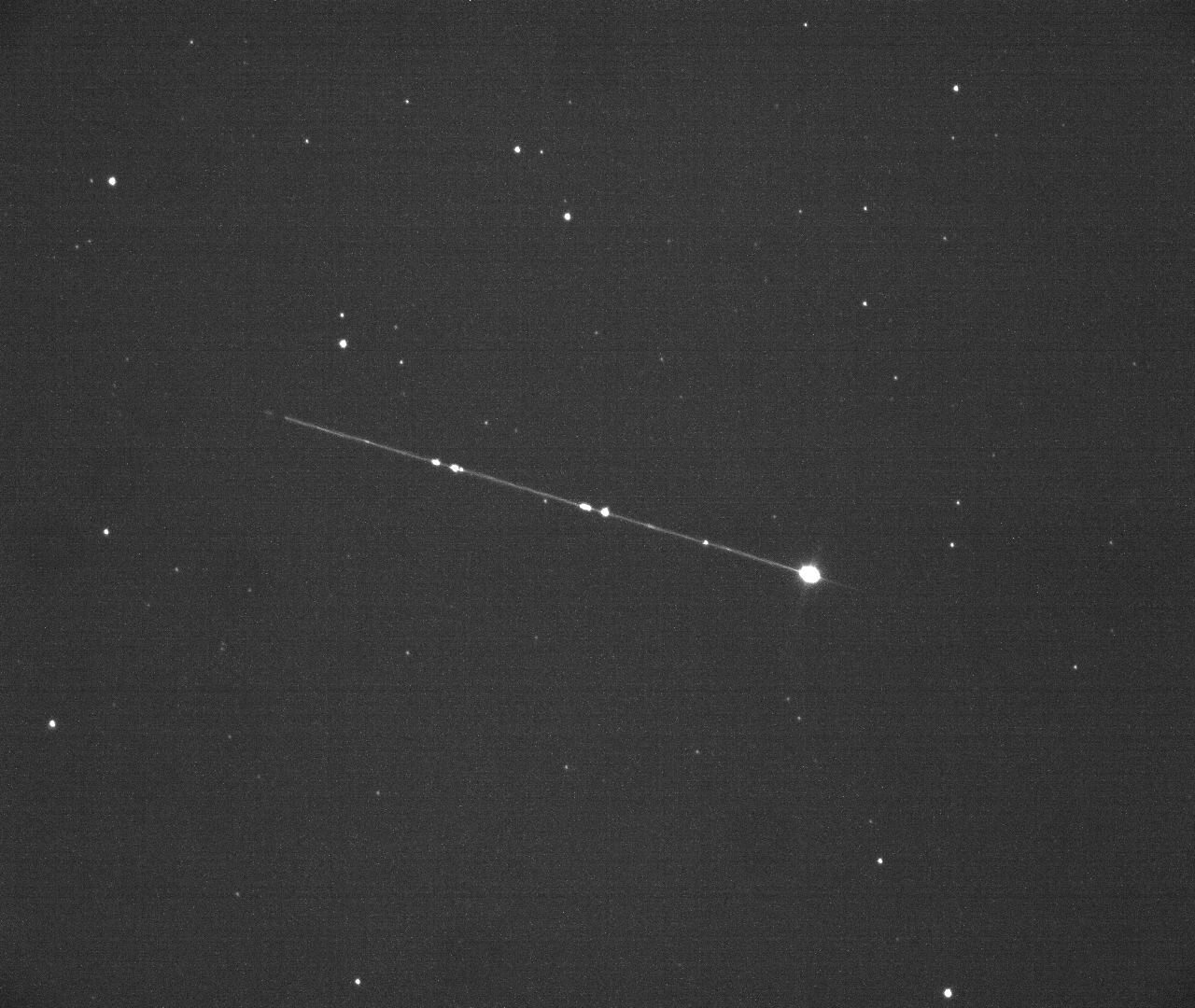}
   \caption{Picture taken of Ajisai (Sat.-no. 16908) on February 11$^\mathrm{th}$, 2014, at 18:58:10 UTC, with an exposure time of 0.6 s. }
   \label{ajisai}
 \end{figure}

\begin{figure}[t]
   \centering
   \includegraphics[width=\columnwidth]{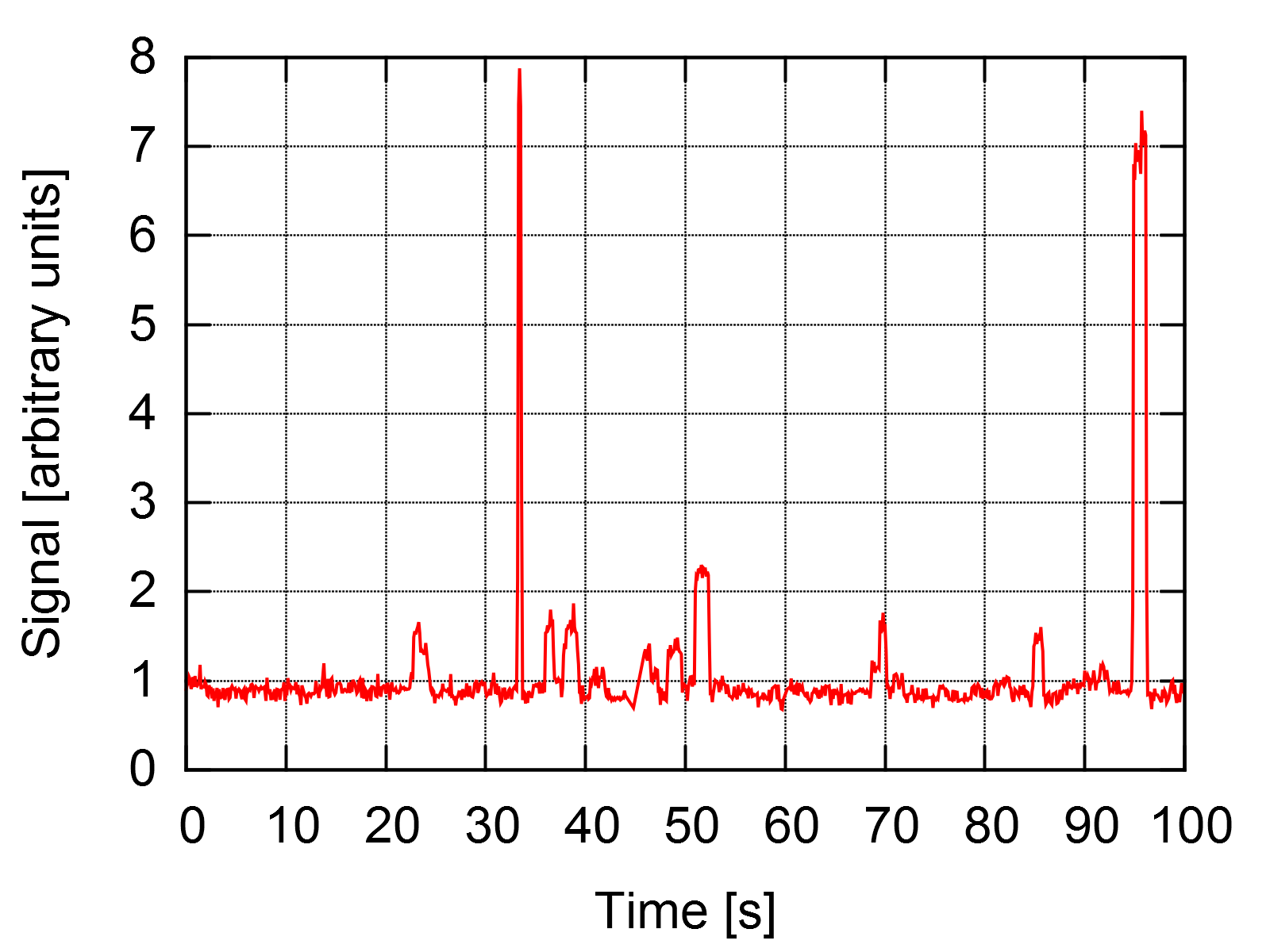}
   \caption{Brightness measured during the transit of a Russian upper stage (SL-8, sat.-no. 21938) at June 25$^\mathrm{th}$, 2014, around 22:40 UTC. The temporal resolution is $\unit[0.1]{s}$.}
   \label{lightcurve}
 \end{figure}

Many LEO objects exhibit considerable brightness variations, often on very short time scales. An example is shown in figure \ref{ajisai}, where clear spots of increased brightness can be found during the $\unit[0.6]{s}$ exposure of an Ajisai transit. A more quantitative measurement is shown in figure \ref{lightcurve}, where the light curve of a Russian upper stage is shown. The light curve has been recorded with continuous tracking and the Zyla camera working at $\unit[10]{Hz}$ (0.1 s exposure time per image).

The origin of these brightness variations are changes of the object's orientation relative to the observer. Specular reflexes of sunlight towards the observer cause bright flashes. If the changing viewing angles are taken into account, such light curves can yield information about the object's rotational movement. Such an evaluation is however not (yet) part of this project.

\subsection{Parallax measurements}
One possible way to determine directly the 3D position of an object is a parallax measurement. For objects in LEO, i.e.\ with distances of up to a few thousand kilometres, a baseline of a few kilometres already yields useful results. Assuming a resolution of two arcseconds in the angular measurement, a baseline of $\unit[100]{km}$, and an object distance of about $\unit[1000]{km}$, the distance can be determined with a resolution of better than $\unit[100]{m}$. For longer baselines the resolution improves almost linearly. However, for distances much larger than $\unit[100]{km}$, the chances for simultaneous visibility of a LEO object decrease seriously.

The main challenge for parallax measurements is the strictly simultaneous exposure of the two cameras at the two viewing locations. With our two Andor cameras, one at main telescope at the observatory and the other one on the portable mount, first simultaneous measurements have been conducted successfully. Currently, campaigns for observations with longer baselines are being planned.


\subsection{Detection of new objects}

\begin{figure}[t]
   \centering
   \includegraphics[width=\columnwidth]{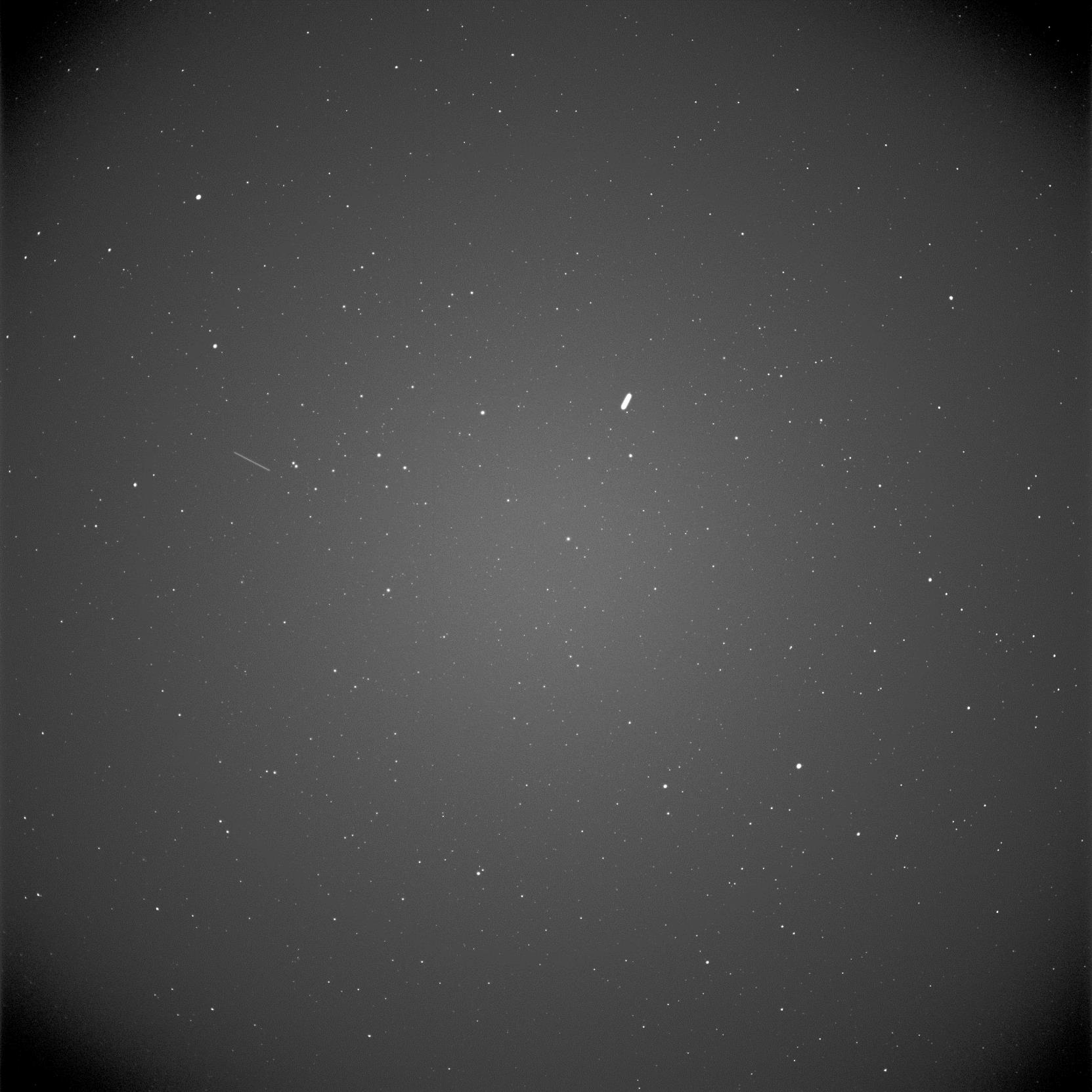}
   \caption{This sample frame from the detection camera shows two objects passing the FOV during one exposure. An automatic image analysis routine detects these tracks and correlates occurrences in subsequent images. The image was taken with the FLI camera and the $\unit[135]{mm}$, $f/2$ photo lens, the FOV is $15^\circ \times 15^\circ$.}
   \label{staring_image}
 \end{figure}

To detect new objects, i.e.\ without prior trajectory information, a camera with a wide-angle lens is used (typically about 5$^\circ$ to 20$^\circ$ field of view). Due to its large resolution, and the relatively low demands on the timing, the FLI ProLine camera is used for this. The mount is fixed or in sidereal tracking, and images are taken continuously, only interrupted by the time needed for the image readout. Typical values are $\unit[1]{s}$ exposure and $\unit[2]{s}$ gap to the next exposure.

Figure \ref{staring_image} shows an image recorded in this way. Due to the large field of view, even fast objects commonly occur in two or more subsequent images and can therefore be distinguished efficiently from noise events. The length of the track and the translation between two images can be used to derive a short-term prediction and move the main tracking telescope to appropriate coordinates. With this telescope, the object's trajectory can be determined accurately.

The detection rate is the product of the field of view and the detection efficiency. The detection efficiency, in turn, is governed by the time the object's light is collected on each single pixel. The larger the magnification, the shorter is the time per pixel, down to a few milliseconds per pixel for fast objects in the main telescope. Simulations with ESA's PROOF software\footnote{\url{http://www.master-model.de/}} \cite{Krag2000687} show that a $\unit[135]{mm}$ photo lens with $f/2$ offers a reasonably high detection efficiency, actually about 50\% higher than our main telescope. The expected detection rate with this set-up is about 20 objects per hour, compared to less than one per hour in the main telescope. First long-time staring mode measurements confirm these numbers.

\section{Active optical measurements}
\label{active}

\subsection{Context}
In our context, active optical measurements denote the illumination of the orbital object with a pulsed laser beam, and the measurement of the object's distance via the light's time of flight. An active optical space debris monitoring system consists of a laser source, a transmitter which collimates the beam and directs it onto the object, and a receiver channel with a single-photon detector and appropriate optics to focus the returning light onto the detector (usually a telescope). Four different configurations are conceivable: 
\begin{enumerate}
\item One single telescope is used as transmitter and receiver (true monostatic)
\item Transmitter and receiver channels use different optical paths and telescopes, but the same mount (bi-static, piggy-back configuration)
\item Transmitter and receiver channels use different optical paths, telescope and mounts. The two mounts are located on the same site, usually a few metres apart from each other (true bi-static)
\item Transmitter and receiver are located at different sites, many kilometres apart from each other (bi-static, long baseline)
\end{enumerate}

Similar systems are in use by the International Laser Ranging Service (ILRS\footnote{http://ilrs.gsfc.nasa.gov/}) for geodetic and other scientific tasks. Compared to this satellite laser ranging (SLR), where all targets are equipped with retroreflectors, space debris observations require significantly higher pulse energies due to the smaller effective optical cross section. While SLR observations are usually conducted with pulse energies of some $\unit[10]{\upmu J}$, even large debris objects require pulse energies of at least a few mJ at $\unit[1]{kHz}$ rate (or even more energy at lower repetition rates). Furthermore, debris observations  require a passive-optical guiding of the observation telescopes since the accuracy of trajectory predictions is usually too low for blind tracking \cite{2013AdSpR..51...21K}.

While SLR observations are conducted routinely for some decades now, laser ranging to space debris objects is a rather new field. EOS Technologies Inc.\ developed a laser ranging system at Mt Stromlo Observatory near Canberra, Australia, and is using it to demonstrate the capability of this technique \cite{smith2006, sang2013}. NASA currently envisages to set-up a sophisticated system which allows for simultaneous distance and velocity measurements using doublet pulses \cite{prasad2013}.
In Shanghai, China, a space debris laser ranging system has gone online in 2008 for testing and demonstration \cite{2012RAA....12..212Z}.
At the Métrologie Optique telescope at the Côte d’Azur in France, a joint team of local researchers and Astrium has installed a laser ranging system and has seen first echoes from debris objects in March 2012 \cite{doi:10.1117/12.2015365}.
Around the same time, the German Aerospace Center (DLR) and the Austrian Space Research Institute have upgraded the Graz (Austria) SLR station with a stronger laser and measured distances to  several dozen debris objects with radar cross sections down to $\unit[0.3]{m^2}$ \cite{2013AdSpR..51...21K}. Currently, DLR is upgrading its own orbital debris research observatory in Stuttgart with a laser ranging channel, while at the same time pursuing further joint experiments and campaigns with the SLR observatories in Graz and Wettzell (Germany). 

\subsection{Upgrade of the orbital debris research observatory}

In a first phase the upgrade, a Nd-YAG laser\footnote{Innolas AOT YAG, \url{http://www.innolas-laser.com/}} with $\unit[300]{\upmu J}$ pulse energy and a repetition rate of $\unit[1]{kHz}$ is used. It will be operated at its fundamental wavelength of $\unit[1064]{nm}$, which promises about one order of magnitude improvement in photon budget compared to the $\unit[532]{nm}$ commonly used in SLR operations \cite{uwe2013}. The configuration will be bi-static using a piggy-back mounting. To reduce the strain on the telescope mount, it is planned to keep the laser stationary on the ground and lead the light onto the transmitter optics via optical fibre. While this approach can probably not be extended to the much higher pulse energies needed for the observation of smaller debris objects, it offers a lean and cost effective way to set up a system for laser ranging at moderate pulse energies. This laser transmitter is expected to offer sufficient power to see echoes from objects equipped with retro-reflectors and possibly very large un-cooperative objects.

In a second phase, it is planned to install a separate laser transmitter with Coudé path, which will allow the use of powerful lasers with up to some $\unit[100]{mJ}$ pulse energies (true bi-static configuration). 

\begin{figure}[t]
   \centering
   \includegraphics[width=\columnwidth]{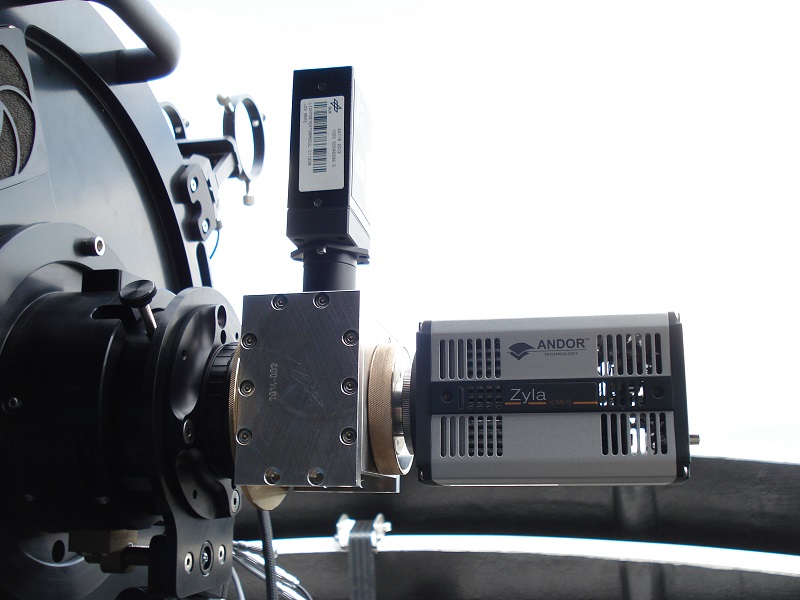}
   \caption{One of the tested single photon detectors mounted on the side-view port of the telescope. Spectral filters and mirrors in the port guide the visible light onto the camera and the infrared photons onto the detector.}
   \label{receiver_channel}
 \end{figure}

For both cases, the receiver channel will be incorporated into the current main telescope's light path. A shortpass filter in front of the camera reflects photons of $\unit[1064]{nm}$ back and via some extra mirrors and lenses onto a single photon detector. Several detectors with a relatively large sensitive area, a low dark count rate and high quantum efficiency are available for this task. Figure \ref{receiver_channel} shows one of the tested detectors on the telescope.

For measuring the timestamps of the incoming photons, both a PicoQuant\footnote{\url{http://www.picoquant.com}} PicoHarp 300 event timer and a White Rabbit\footnote{\url{http://www.ohwr.org/projects/white-rabbit}} FMC-DEL card are available, both with sub-nanosecond accuracy. The FMC-DEL card additionally offers an interface to UTC time via GPS, and the possibility to produce custom gate pulses for the detector. Assuming a laser repetition rate of $\unit[1]{kHz}$ and a distance search window of $\unit[1]{km}$ ($\overset{\scriptscriptstyle\wedge}{=}   \unit[6]{\upmu s}$), the duty cycle of the detector can be kept below 1\%, thus reducing the recorded noise.

\begin{figure}[t]
   \centering
   \includegraphics[width=\columnwidth]{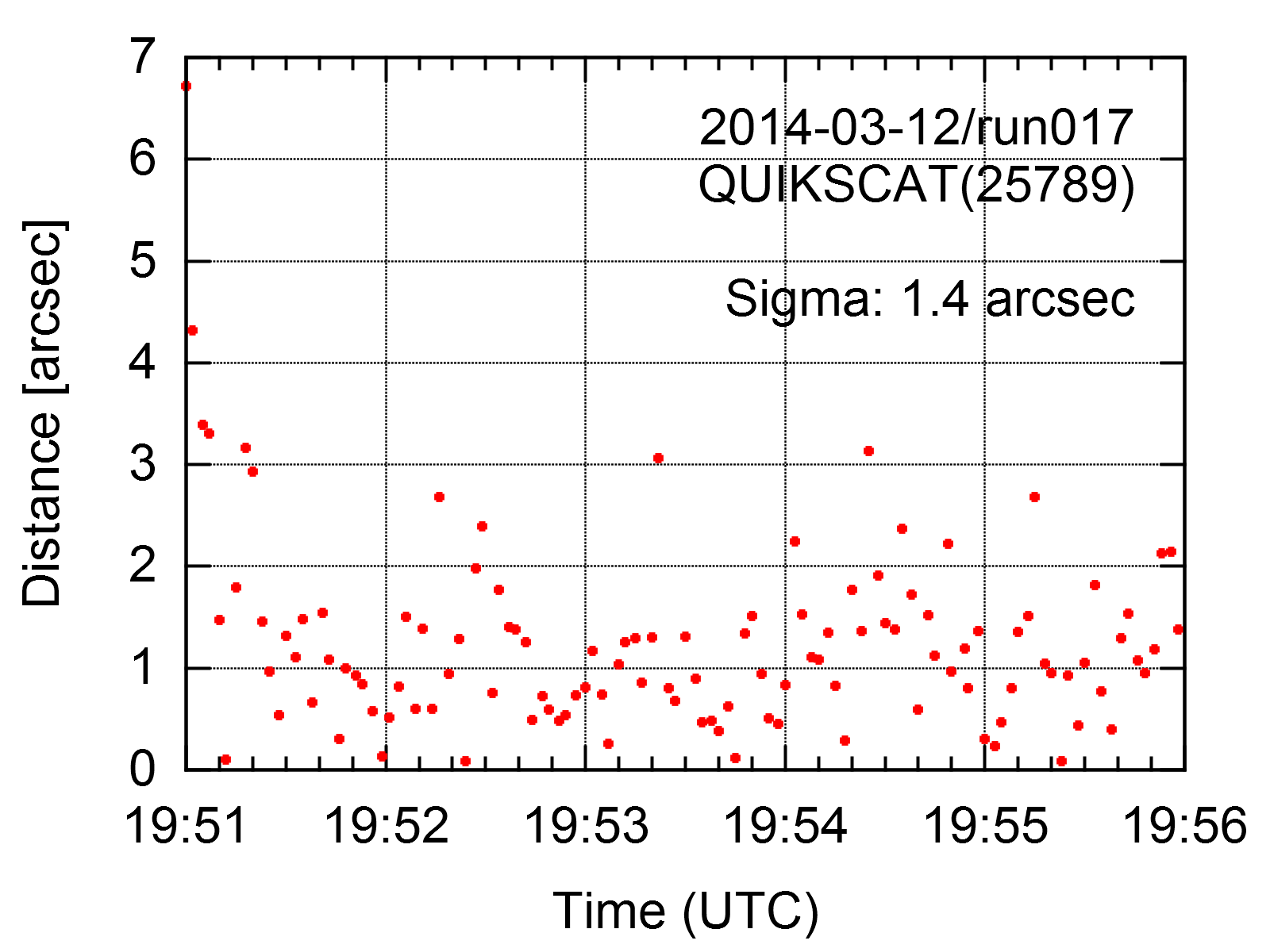}
   \caption{Tracking accuracy example for transit of object QuikSCAT (Sat.-no. 25789) on March 12$^\mathrm{th}$, 2014. The vertical axis shows the deviation of the object measured from its target point in the camera.}
   \label{tracking}
 \end{figure}

For successful laser ranging, the accurate tracking of fast LEO objects is a necessary precondition. Using a closed loop control system, the telescope currently achieves a tracking accuracy of about two arcseconds ($1\sigma$) on most objects. One example if shown in Figure \ref{tracking}, where the object could be tracked with an accuracy of $\sigma = 1.4"$ for several minutes after an initial acquisition phase. This is close to the seeing limit and should be sufficient to keep the laser beam on target.

\subsection{Joint experiments between Stuttgart, Wettzell and Graz}
As part of an ESA study, DLR, the Technical University of Munich and the Austrian Space Research Institute will install a $\unit[100]{mJ}$ laser ranging system at the SLR station in Wettzell. In this project, long baseline bi-static ranging measurements are planned, using Wettzell as transmitter and Stuttgart and Graz as secondary receivers. First successful tests of such bi-static measurements have been conducted using Graz as transmitter and the SLR stations Wettzell, Zimmerwald and Herstmonceux as receivers \cite{Kirchner2013}. During the new project, several objects will be observed during the course of at least one year and the possible improvement of the trajectory prediction due to laser ranging will be evaluated quantitatively.

\section{Conclusion}
The orbital debris research observatory in Stuttgart is used to develop, evaluate and demonstrate optical methods for observation of objects in LEO. Passive-optical observations are conducted since April 2013, using different cameras and optical systems. It could be shown that the passive-optical approach offers great potential both for detection of new objects and the accurate measurement of the object's position (angles only). However, the distance to the object can only be inferred by indirect means and comparatively low accuracy.

One possibility to measure the distance to an orbital object directly is the use of laser ranging. Currently, the orbital debris research station is being upgraded with a laser transmitter and receiver channel. First ranging tests to orbital objects are planned for 2015. Additionally, the German Aerospace Center works in close cooperation with nearby SLR stations, e.g.\ Graz and Wettzell, to test and evaluate novel approaches like the long-baseline bi-static ranging.

\bibliography{references}

\providecommand{\bysame}{\leavevmode\hbox to3em{\hrulefill}\thinspace}
\providecommand{\MR}{\relax\ifhmode\unskip\space\fi MR }
\providecommand{\MRhref}[2]{%
  \href{http://www.ams.org/mathscinet-getitem?mr=#1}{#2}
}
\providecommand{\href}[2]{#2}
\begin{thebibliography}{10}

\bibitem{andor}
{Andor}, \emph{{iXon Ultra 897}},
  \url{http://www.andor.com/pdfs/specifications/Andor_iXon_Ultra_897_Specifications.pdf},
  [Online; accessed 26-June-2013].

\bibitem{andor_zyla}
\bysame, \emph{{Zyla}},
  \url{http://www.andor.com/pdfs/specifications/Andor_Zyla_sCMOS_Specifications.pdf},
  [Online; accessed 04-November-2013].

\bibitem{fli}
{Finger Lakes Instruments}, \emph{{ProLine 16083 data sheet}},
  \url{http://www.flicamera.com/spec_sheets/PL16803.pdf}, [Online; accessed
  26-June-2013].

\bibitem{Hampf2013}
D.~{Hampf}, W.~{Riede}, G.~{Stöckle}, and I.~{Buske}, \emph{{Ground-Based
  Optical Position Measurements of Space Debris in Low Earth Orbits}},
  Deutscher Luft- und Raumfahrtkongress, 2013.

\bibitem{doi:10.1117/12.2015365}
L.~Hennegrave, M.~Pyanet, H.~Haag, G.~Blanchet, B.~Esmiller, S.~Vial,
  E.~Samain, J.~Paris, and D.~Albanese, \emph{{Laser ranging with the MéO
  telescope to improve orbital accuracy of space debris}}, Proc. SPIE, vol.
  8739, 2013, pp.~87390J--87390J--12.

\bibitem{2013AdSpR..51...21K}
G.~{Kirchner}, F.~{Koidl}, F.~{Friederich}, I.~{Buske}, U.~{V{\"o}lker}, and
  W.~{Riede}, \emph{{Laser measurements to space debris from Graz SLR
  station}}, Advances in Space Research \textbf{51} (2013), 21--24.

\bibitem{Kirchner2013}
G.~{Kirchner}, F.~{Koidl}, M.~{Ploner}, et~al., \emph{{Multistatic Laser
  Ranging to Space Debris}}, 18th International Workshop on Laser Ranging
  \textbf{13-0213} (2013).

\bibitem{Krag2000687}
H.~Krag, P.~Beltrami-Karlezi, J.~Bendisch, H.~Klinkrad, D.~Rex, J.~Rosebrock,
  and T.~Schildknecht, \emph{{PROOF — The extension of ESA's MASTER Model to
  predict debris detections}}, Acta Astronautica \textbf{47} (2000), no.~2–9,
  687 -- 697.

\bibitem{2013EUCAS...4..735L}
J.-C. {Liou}, \emph{{Engineering and technology challenges for active debris
  removal}}, EUCASS Proceedings Series \textbf{4} (2013), 735--748.

\bibitem{2013JGCD...36..743M}
J.~{Missel} and D.~{Mortari}, \emph{{Removing Space Debris Through Sequential
  Captures and Ejections}}, Journal of Guidance Control Dynamics \textbf{36}
  (2013), 743--752.

\bibitem{quarterlynews2014i1}
{National Aeronautics and Space Administration}, \emph{{Orbital Debris
  Quarterly News}},  \textbf{1} (2014), [Online; accessed 29-July-2014].

\bibitem{2014AcAau..93..418P}
C.~R. {Phipps}, \emph{{A laser-optical system to re-enter or lower low Earth
  orbit space debris}}, Acta Astronautica \textbf{93} (2014), 418--429.

\bibitem{prasad2013}
N.~{Prasad}, A.~{DiMarcantonio}, and A.~{Van Rudd}, \emph{{Development of
  Coherent Laser Radar for Space Situational Awareness Applications}}, AMOS
  Conf. Techn. Papers (2013).

\bibitem{jonas}
J.~{Radtke}, \emph{{Visibility studies and orbit determination of space debris
  objects for a combined passive-optical and laser ranging station}}, 2012,
  Diploma thesis at Institute for Technical Physics of DLR, Stuttgart.

\bibitem{sang2013}
J.~{Sang}, I.~{Ritchie}, M.~{Pearson}, and C.~{Smith}, \emph{{Results and
  Analyses of Debris Tracking from Mt Stromlo}}, AMOS Conf. Techn. Papers
  (2013).

\bibitem{smith2006}
C.~{Smith}, \emph{{The EOS Space Debris Tracking System}}, AMOS Conf. Techn.
  Papers (2006).

\bibitem{UNO_space_debris_mitigation}
{United Nations Office for Outer Space Affairs}, \emph{{Space Debris Mitigation
  Guidelines}},  (2010).

\bibitem{uwe2013}
U.~{V{\"o}lker}, F.~{Friederich}, I.~{Buske}, D.~{Hampf}, W.~{Riede}, and
  A.~{Giesen}, \emph{Laser based observation of space debris: Taking benefits
  from the fundamental wave}, Proc. of 6th european conference on space debris,
  2013.

\bibitem{2012RAA....12..212Z}
Z.-P. {Zhang}, F.-M. {Yang}, H.-F. {Zhang}, Z.-B. {Wu}, J.-P. {Chen}, P.~{Li},
  and W.-D. {Meng}, \emph{{The use of laser ranging to measure space debris}},
  Research in Astronomy and Astrophysics \textbf{12} (2012), 212--218.

\end{thebibliography}
\bibliographystyle{plain}

\end{document}